\documentclass[aps,prd,twocolumn,letterpaper,superscriptaddress, showpacs, amsmath, amssymb]{revtex4}

\usepackage{graphicx,color}
\usepackage{graphicx}
\usepackage{amsmath,amssymb}
\usepackage{url}
\newcommand{\src}{M\thinspace 31}
\newcommand{\chandra}{Chandra}
\newcommand{\xmm}{XMM-Newton}
\newcommand{\suzaku}{Suzaku}

\newcommand{\mdm}{$M_{DM}^{FOV}$}
\newcommand{\mixing}{${\rm sin}^2 \theta$}
\newcommand{\arcmin}{$^\prime$}
\newcommand{\msun}{$M_\odot$}
\newcommand{\mparticle}{$m_s$}

\begin{document}


\title{Sterile neutrino dark matter bounds from galaxies of the Local Group}

\author{Shunsaku Horiuchi}
\email{s.horiuchi@uci.edu}
\affiliation{Center for Cosmology, Department of Physics and Astronomy, 4129 Frederick Reines Hall, University of California, Irvine, CA 92697-4575}
\author{Philip J.~Humphrey}
\affiliation{Center for Cosmology, Department of Physics and Astronomy, 4129 Frederick Reines Hall, University of California, Irvine, CA 92697-4575}
\author{Jose O\~norbe}
\affiliation{Center for Cosmology, Department of Physics and Astronomy, 4129 Frederick Reines Hall, University of California, Irvine, CA 92697-4575}
\author{Kevork N.~Abazajian}
\affiliation{Center for Cosmology, Department of Physics and Astronomy, 4129 Frederick Reines Hall, University of California, Irvine, CA 92697-4575}
\author{Manoj Kaplinghat}
\affiliation{Center for Cosmology, Department of Physics and Astronomy, 4129 Frederick Reines Hall, University of California, Irvine, CA 92697-4575}
\author{Shea Garrison-Kimmel}
\affiliation{Center for Cosmology, Department of Physics and Astronomy, 4129 Frederick Reines Hall, University of California, Irvine, CA 92697-4575}

\date{date}

\begin{abstract}
We show that the canonical oscillation-based (non-resonant) production of sterile neutrino dark matter is inconsistent at $>99$\% confidence with observations of galaxies in the Local Group. We set lower limits on the non-resonant sterile neutrino mass of $2.5$ keV (equivalent to $0.7$ keV thermal mass) using phase-space densities derived for dwarf satellite galaxies of the Milky Way, as well as limits of $8.8$ keV (equivalent to $1.8$ keV thermal mass) based on subhalo counts of $N$-body simulations of M 31 analogues. Combined with improved upper mass limits derived from significantly deeper X-ray data of M 31 with full consideration for background variations, we show that there remains little room for non-resonant production if sterile neutrinos are to explain $100$\% of the dark matter abundance. Resonant and non-oscillation sterile neutrino production remain viable mechanisms for generating sufficient dark matter sterile neutrinos.
\end{abstract}


\maketitle


\section{ Introduction}
The particle nature of dark matter (DM) is among the most intriguing questions in modern physics, and many extensions to the Standard Model of particle physics have been considered \cite{Jungman1995df,Bertone2004pz, Feng2010gw}. Among the highly motivated candidates is the sterile (singlet) neutrino with a mass in the keV range \cite{Boyarsky2009ix,Kusenko2009up}. In the simplest original Dodelson-Widrow scenario \cite{Dodelson1993je}, they are produced in the early Universe via oscillations with active neutrinos which are non-resonant in the presence of negligible lepton asymmetry (DW; see also \citep{Dolgov2000ew,Abazajian2001nj}). Production via resonant oscillations in the presence of a lepton asymmetry (resonant production, RP \cite{Shi1998km,Abazajian2001nj,Laine2008pg}), via interactions with the inflaton \cite{Kusenko2006rh,Shaposhnikov2006xi}, and scalar production \cite{Merle2013wta} have also been proposed. In addition to DM, sterile neutrinos may also explain the observed velocities of pulsars \cite{Kusenko1997sp,Kusenko2006rh}.

The DW sterile neutrino is warm dark matter (WDM) with a non-negligible velocity dispersion. This suppresses the matter power spectrum below the free-streaming scale and affects DM structures. Conversely, information of the matter power spectrum on small scales can be used to constrain the sterile neutrino properties. By modeling the SDSS Lyman-$\alpha$ forest flux power spectrum, lower limits of $m_s^{\rm DW} \gtrsim 13$~keV have been found \cite{Abazajian2005xn,Viel2006kd,Seljak2006qw,Boyarsky2008xj}; with recent Keck data, $m_s^{\rm DW} \gtrsim 22$~keV at $2\sigma$ \cite{Viel2013fqw}. Lower limits of $m_s^{\rm DW} \gtrsim 13$~keV have also been placed by requiring the number of subhalos in $N$-body simulations to be larger than the number of observed dwarf spheroidal galaxies (dSphs) of the Milky Way (MW) \cite{Polisensky2010rw}. Limits have also been placed using high-$z$ observations of gamma-ray bursts \cite{deSouza2013wsa} and galaxies \cite{Abazajian2013aa}.

At the same time, sterile neutrinos are not completely stable and their radiative decays into active neutrinos provides a compelling search opportunity \cite{Abazajian2001vt}. Due to detector capabilities, current X-ray searches probe masses of a few keV and above. Many DM sources have been studied, ranging from the X-ray background, galaxy clusters, nearby galaxies, and our own MW (see, e.g., Ref.~\cite{Boyarsky2006fg} and references therein). The M $31$ galaxy yields some of the strongest constraints, and for the DW sterile neutrino, previous works have limited the mass to $m_s^{\rm DW} \lesssim 3$ keV \cite{Watson2006qb,Boyarsky2007ay,Watson2011dw}.

When combined, the lower and upper limits seemingly already rule out the DW sterile neutrino. However, modeling the Lyman-$\alpha$ forest flux requires hydrodynamic simulations with implicit assumptions about the thermal history of the absorbing gas and its ionizing background. When these assumptions are relaxed, the mass limits are diluted (see, e.g., Figures 11 and 12 of Ref.~\cite{Viel2013fqw} and discussions therein). The comparison of subhalos to MW dSphs assumes a factor $\sim 4$ correction for the number of dSphs being missed by current surveys; without the correction, limits are weakened. Although dSphs are no doubt being missed, this introduces a large uncertainty in the limit. Given these large systematic uncertainties, additional constraints are required to make definitive conclusions regarding the viability of the DW mechanism. 

More robust lower limits have been placed by exploiting the limited phase-space packing of sterile neutrinos \cite{Tremaine1979we,Dalcanton2000hn}. For a given primordial momentum distribution of sterile neutrinos, a theoretical maximum phase-space density exists. Comparing these to the phase-space densities estimated from MW dSphs, limits of $m_s^{\rm DW} \gtrsim 1.8$~keV have been set \cite{Gorbunov2008ka,Boyarsky2008ju}, leaving a small window for the DW sterile neutrino production mechanism to generate $100$\% of the observed DM abundance \cite{Boyarsky2008ju}. 

In this paper, we revisit the lower and upper limits on sterile neutrinos and address the viability of the DW mechanism in explaining $100$\% of the observed DM abundance. We improve both lower limits placed from phase-space arguments and subhalo counts, as well as X-ray upper limits. For phase-space limits, we consider new MW dSphs not considered in previous works, and we also address the main uncertainty in estimating the DM velocity dispersion. For subhalo counts, we focus on M 31 and do not rely on uncertain incompleteness corrections. Finally, for X-ray constraints we use the largest and deepest data assembled of M $31$ and include full background uncertainties. Using our improved and more robust constraints, we are able to rule out the DW sterile neutrino as a viable DM candidate at more than $99$\% confidence level (C.L.). 

In Section \ref{sec:phasespace} we discuss our new phase-space constraints, followed by subhalo count limits in Section \ref{sec:counts}. We discuss X-ray constraints in Section \ref{sec:xray} and conclude in Section \ref{sec:conclusion}. Throughout, as our focus is on the DM sterile neutrino we opt to use the mass $m_s^{\rm DW}$ as our main parameter. However, it is also common to quote the mass of a thermal WDM particle, $m_{\rm WDM}$, which is related to $m_s^{\rm DW} $ by $m_s^{\rm DW} \approx 4.379 \,{\rm keV} (m_{\rm WDM}/1 \, {\rm keV})^{4/3} (\Omega_m/0.238)^{-1/3} (h/0.73)^{-2/3} $ \cite{Viel2005qj}. 

\section{Phase-space density limits} \label{sec:phasespace}

\subsection{General considerations}

Lower limits on the DM particle mass are based on Liouville's theorem. For dissipationless and collisionless particles, the phase-space density cannot increase, and its maximum does not change with time. Estimates of the coarse-grained phase-space density made using astrophysical observations must therefore satisfy $Q < q_{\rm max}$, where $Q$ is the coarse-grained phase-space density and $q_{\rm max}$ is the maximal fine-grained phase-space density \cite{Hogan2000bv}. Since $q_{\rm max}$ depends on the primordial DM properties, the inequality can be used to limit, e.g., the DM mass. 

The momentum distribution of DW sterile neutrinos at production is well approximated by $f_s(p) = \beta (e^{p/T}+1)^{-1}$. Here, $p$ is momentum and $T$ is temperature. If $\beta = 1$, one recovers the thermal Fermi-Dirac distribution, and the fine-grained phase-space density maximum is \cite{Boyarsky2008ju},
\begin{eqnarray}
q_{\rm max}^{\rm FD} &=& \frac{g \, m^4}{ 2 (2 \pi \hbar)^3} \\ \nonumber
&\approx& 5 \times 10^{-4} 
\left( \frac{g}{2} \right) 
\left( \frac{m}{1 \, {\rm keV}} \right)^{4} 
{\rm M_\odot pc^{-3}} ({\rm km /s})^{-3},
\end{eqnarray}
where $g$ is the number of spin degrees of freedom. For the DW sterile neutrino, $\beta$ can be set by the requirement to obtain the correct relic density, $\beta \approx \Omega_{dm} h^2 (m_s^{\rm DW}/94 \, {\rm eV})^{-1}$ \cite{Dodelson1993je}, so that,
\begin{eqnarray} \label{eq:Qmax}
q_{\rm max}^{\rm DW} &=& \eta\frac{\beta g \, m^4}{ 2 (2 \pi \hbar)^3} \\ \nonumber
&\approx& 6 \times 10^{-6} 
\left( \frac{\eta}{1.25} \right) \left( \frac{g}{2} \right) 
\left( \frac{m_s^{\rm DW}}{1 \, {\rm keV}} \right)^{3} 
\left( \frac{\Omega_{dm} h^2}{0.105} \right) \\ \nonumber
&& {\rm M_\odot pc^{-3}} ({\rm km /s})^{-3}.
\end{eqnarray}
Here, the additional $\eta$ factor is a correction factor due to the fact that $\beta$ is not strictly a constant. The $\beta =$ constant estimate is only valid for $T\lesssim 200$ MeV where the number of particle degrees of freedom can be taken to be constant; above this, the activation of the extra gluon and quark degrees of freedom requires a numerical treatment \cite{Abazajian2005gj,Asaka2006nq,Asaka2006rw}. The earlier production momenta has the effect of shifting the momentum distribution colder (see, e.g., Figure 1 of Ref.~\cite{Abazajian2005gj}), implying larger fine-grained phase-space maxima. We find the effect to be a $\sim 20$--$25$\% increase over a wide sterile neutrino mass range, and conservatively adopt $\eta = 1.25$.

\begin{table*}
\caption{Column (1): name, (2): distance, (3): stellar dispersion, (4): half-light radius, (5): the total mass within $r_h$; all from Ref.~\cite{Wolf2009tu}. Column (6): Q values [in units of $10^{-5} ({\rm M_\odot/ pc^3 })({\rm km/s})^{-3}$] estimated with Eq.~(\ref{eq:Qestimate}) using columns (3) and (4). Column (7): number of matched subhalos in VL2 used to obtain columns (8) and (9). Column (8): $\bar{Q}$ values [in units of $10^{-5} ({\rm M_\odot/ pc^3 })({\rm km/s})^{-3}$] estimated using NFW profile. Column (9): $\bar{Q}$ values [in units of $10^{-5} ({\rm M_\odot/ pc^3 })({\rm km/s})^{-3}$] estimated using pseudo-isothermal profile. Column (10): lower mass $m_s^{\rm DW}$ estimated using column (9).
\label{tab:QandLimits}}
\begin{ruledtabular}
\begin{tabular}{lrrrrrrrrr}
name			& $d$ [kpc]	& $\sigma_*$ [km/s]	& $r_h$ [pc]		& $M_h$ [$10^6 M_\odot$]	& $Q_{\rm MB}(\eta_*=1)$	& $N_{\rm sh}$ 	
	& $\bar{Q}_{\rm sim}^{\rm NFW}$	&$\bar{Q}_{\rm sim}^{\rm Iso}$		& $m_s^{\rm DW}$ [keV]		\\
(1) 	& (2)		& (3)		& (4)		& (5)		& (6)		& (7)		& (8)		& (9)		& (10)	\\	
\hline 
Draco			& $76\pm5$		& $10.1\pm 0.5$	& $291^{+14}_{-14}$	& $2.11^{+3.1}_{-3.1} $	& $1.2 \pm0.1 $	& $3$	& $0.53\pm0.15$	& $0.59\pm0.07$	& $>1.1\pm0.04$	\\
Carina			& $105\pm2$		& $6.4\pm 0.2$		& $334^{+37}_{-37}$	& $9.56^{+0.95}_{-0.90}$	& $1.5 \pm0.3 $	& $8$	& $0.67\pm0.14$	& $0.66\pm0.06$	& $>1.1\pm0.03$	\\
Hercules			& $133\pm6$		& $5.1\pm 0.9$		& $305^{+26}_{-26}$	& $7.50^{+5.72}_{-3.14}$	& $2.2 \pm0.5 $	& $37$	& $0.75\pm0.21$	& $0.71\pm0.23$	& $>1.0\pm0.1$	\\
Leo II			& $233\pm15$		& $6.6\pm 0.5$		& $233^{+17}_{-17}$	& $7.25^{+1.19}_{-1.01}$	& $3.0 \pm0.5 $	& $30$	& $1.1\pm0.3$		& $1.2\pm0.3$		& $>1.2\pm0.1$	\\
Ursa Major II		& $32\pm4$		& $6.7\pm 1.4$		& $184^{+33}_{-33}$	& $7.91^{+5.59}_{-3.14}$	& $4.7 \pm1.9 $	& $8$	& $1.1\pm0.4$		& $1.1\pm0.4$		& $>1.2\pm0.2$	\\
Leo T			& $407\pm38$		& $7.8\pm 1.6$		& $152^{+21}_{-21}$	& $7.37^{+4.84}_{-2.96}$	& $5.9 \pm2.0 $	& $33$	& $1.4\pm0.4$		& $2.0\pm0.7$		& $>1.5\pm0.2$	\\
Leo IV			& $160\pm15$		& $3.3\pm 1.7$		& $151^{+34}_{-44}$	& $1.14^{+3.50}_{-0.92}$	& $14 \pm11 $		& $80$	& $2.4\pm0.8$		& $1.6\pm0.8$		& $>1.4\pm0.2$	 \\
Canes Venativi II	& $160\pm5$		& $4.6\pm 1.0$		& $97^{+18}_{-13}$		& $1.43^{+1.01}_{-0.59}$	& $24 \pm10 $		& $18$	& $4.6\pm1.5$		& $5.3\pm2.7$		& $>2.0\pm0.3$	\\
Coma Berenices	& $44\pm4$		& $4.6\pm 0.8$		& $100^{+13}_{-13}$	& $1.97^{+0.88}_{-0.60}$	& $26 \pm8.7$		& $15$	& $5.2\pm2.4$		& $5.7\pm2.9$		& $>2.1\pm0.4$	\\
\hline 
Segue I			& $23\pm2$		& $4.3\pm 1.1$		& $38^{+10}_{-7}$		& $0.60^{+0.51}_{-0.28}$	& $170 \pm100   $	& $7$	& $30\pm16$		& $36\pm23$		& $>3.9\pm0.8$	\\
\end{tabular}	
\end{ruledtabular}
\end{table*}

The coarse-grained phase-space density $Q$ is defined as the mass density in a \emph{finite} six-dimensional phase-space volume at time $t$. There are multiple definitions in the literature. A popular choice is the pseudo-phase-space density \cite{Hogan2000bv},
\begin{equation}\label{eq:Q1}
Q_{\rm HD00} \equiv \frac{ \bar{\rho}}{(3\sigma^2)^{3/2}},
\end{equation}
where $\bar{\rho}$ is the average DM density and $\sigma$ is the one-dimensional DM velocity dispersion. A more realistic phase-space volume can be defined from adopting a Maxwellian velocity distribution \cite{Shao2012cg},
\begin{equation}\label{eq:Q2}
Q_{\rm MB} \equiv \frac{ \bar{\rho}}{(2 \pi \sigma^2)^{3/2}} \approx 0.33 \, Q_{\rm HD00}, 
\end{equation}
where $(2\pi \sigma^2)^{-3/2}$ is the maximum density in velocity space. Finally, the mass density can be defined very conservatively based on the whole available phase-space volume $\Delta { x}\, \Delta{ v}=(4\pi/3)^2R^3v_\infty^3$, with $v_\infty=\sqrt{6}\sigma$ \cite{Boyarsky2008ju},
\begin{equation}\label{eq:Q3}
Q_{\rm Boy} \equiv \frac{ \bar{\rho} }{ (8 \pi \sqrt{6} \sigma^3)} \approx 0.08 \, Q_{\rm HD00}.
\end{equation}
For the rest of the paper, we focus on $Q_{\rm MB}$, but results for other definitions can be easily obtained from the above scaling relations. 

The dSph satellites of the MW provide the optimum locations for estimating the coarse-grained phase-space density \cite{Dalcanton2000hn}, and have been recently investigated by Refs.~\cite{Gorbunov2008ka,Boyarsky2008ju}. We estimate the coarse-grained phase-space assuming that the density is constant within $r_h$,  and use the mass estimator $M_h \approx 3 \sigma_*^2 r_h / 3$ \cite{Wolf2009tu}. The mean density can then be written $\bar{\rho} = ( 9 \sigma_*^2 ) / (4\pi G r_h^2)$, which yields
\begin{eqnarray} \label{eq:Qestimate}
Q_{\rm MB} &=& \frac{9}{2 (2 \pi)^{5/2} G r_h^2 \eta_*^3 \sigma_*} \\ \nonumber
&\approx& 1.1 \times 10^{-4} 
\eta_*^{-3}
\left( \frac{\sigma_* }{10 \, {\rm km/s}} \right)^{-1} 
\left( \frac{r_h}{100 \, {\rm pc}} \right)^{-2} \\ \nonumber
&& {\rm M_\odot pc^{-3}} ({\rm km /s})^{-3},
\end{eqnarray}
for the Maxwellian phase-space density. 
While simple, this method has a large uncertainty associated with how to estimate the dark matter velocity dispersion, $\sigma$, from the observed stellar velocity dispersion, $\sigma_*$. In previous works, this has been replaced by an ignorance parameter, $\eta_* = \sigma / \sigma_*$, assumed to be of order unity \cite{Gorbunov2008ka,Boyarsky2008ju}. Since the mass limit scales as $Q^{1/3}$, the uncertainty in $\eta_*$ directly affects the limit, $m_s^{\rm DW} \propto 1/\eta_* $. 

Nevertheless, we first estimate $Q_{\rm MB}$ for the MW satellites, adopting values of $\sigma_*$ and $r_h$ from Ref.~\cite{Wolf2009tu}, and assuming $\eta_* = 1$. These are shown in the sixth column of Table \ref{tab:QandLimits}. We stress that these are conservative approximations of the mean phase-space density within $r_h$, rather than the true central densities. The values of $Q_{\rm HD00}$ and $Q_{\rm Boy}$ are obtainable via the scalings Eqs.~(\ref{eq:Q1}--\ref{eq:Q3}). 

The uncertainties on $Q_{\rm MB}$ are derived assuming Gaussian statistics and follow from the uncertainties in the measured $\sigma_*$ and $r_h$ only. However, systematic effects likely dominate the uncertainty. For example, Ursa Major II shows circumstantial evidence of ongoing tidal disruption \cite{Simon2007dq}. Coma Berenices shares some properties with Ursa Major II, although there is no known tidal stream near its position \cite{Simon2007dq} and additional observations are consistent with no ongoing tidal disruption \citep{Munoz2009hj}. At the extreme is Willman I. It has a large estimated $Q_{\rm MB} \approx (1.1 \pm 0.6) \times 10^{-3} ({\rm M_\odot/ pc^3 })({\rm km/s})^{-3}$, but there is compelling evidence of tidal disruption by the MW, and this is likely an overestimate. We therefore omit Willman I from Table \ref{tab:QandLimits}. Segue 1, which also has a high inferred $Q$, is among the faintest dwarfs recently discovered by the SDSS \cite{Belokurov2006ph} and its properties are determined with limited stellar spectroscopy data \cite{Simon2007dq,Simon2010ek}.

\subsection{Phase-space constraints from $N$-body simulations}

In the previous section, the phase-space density was estimated assuming $\eta_* = \sigma / \sigma_*=1$. However, this is not expected to be generally true. Here, we estimate the phase-space density directly from the DM density profiles of subhalos that can host the MW dSphs. For this purpose, we use the subhalos of the Via Lactea II (VL2) simulation \cite{Diemand2008in}. Although this is a $\Lambda$CDM simulation, CDM subhalos are good approximations for WDM subhalos on scales greater than the core radius. For the WDM masses of interest, the core radii are small enough that we are in the CDM-like regime (see also, e.g., Refs.~\cite{VillaescusaNavarro2010qy,Maccio2012qf}). To illustrate this point, however, we will consider both the NFW profile and the pseudo-isothermal profile, and derive $Q$ estimates for both (columns 8 and 9 of Table \ref{tab:QandLimits}).

The NFW profile is a commonly used two-parameter fit to dissipationless $N$-body simulations \cite{Navarro1996gj},
\begin{equation}
\rho_{\rm NFW}(r) = \frac{ \rho_0}{(r/r_s)(1+r/r_s)^2},
\end{equation} 
where $r_s$ and $\rho_0$ can conveniently be written as functions of the maximum circular velocity, $V_{\rm max}$, and the radius at which $V_{\rm max}$ is reached, $R_{\rm max}$. 

The pseudo-isothermal density profile is a good fit to WDM density profiles on scales comparable to the core size, and furthermore gives a good estimate of the phase-space density \cite{Shao2012cg},
\begin{equation}
\rho_{\rm iso}(r) = \frac{ \rho_c}{1+(r/r_c)^2},
\end{equation}
where $\rho_c$ and $r_c$ are the core density and radius. The circular velocity asymptotes to $V_{\rm max} = (4\pi G \rho_0 r_c^2)^{1/2}$, so the profile can be defined by the parameters $(V_{\rm max},r_c)$.

First, values of $(R_{\rm max},V_{\rm max})$ for all VL2 subhalos are obtained. For the NFW profile, these parameters define the density profile. However, there is no well-defined $R_{\rm max}$ for the pseudo-isothermal profile, and $V_{\rm max}$ alone does not define the profile. We therefore also require that the subhalo has the correct mass $M_h$ within the half-light radius to estimate the pseudo-isothermal profile parameters. 

Next, subhalos that can host the MW dSphs are selected. This involves selecting subhalos that have the correct distance to the main host halo and a reasonable $V_{\rm max}$. A tolerance of $\pm 3 \sigma$ is adopted for the MW dSph distances. Subhalos with $V_{\rm max} > 60$ km/s are considered LMC/SMC analogues and are excluded. Subhalos with unreasonably small $V_{\rm max}<10$ km/s are also excluded. For the NFW profile, subhalos must also have the correct mass within $r_h$; once again, a tolerance of $\pm 3 \sigma$ is adopted. The isothermal profile by construction already have the correct mass. The resulting number of subhalos that can host MW dSphs in both the NFW and pseudo-isothermal profiles, $N_{\rm sh}$, is listed in Table \ref{tab:QandLimits}. 

Finally, the phase-space density within $r_h$ is calculated for each subhalo following Eq.~(\ref{eq:Q2}), both for NFW and pseudo-isothermal profiles. The average density is obtained from the profiles, and the DM velocity dispersion is determined for each profile from the spherical Jeans equation assuming an isotropic velocity dispersion tensor. The mean of $N_{sh}$ subhalos, $\bar{Q}_{\rm sim}^{\rm NFW}$ and $\bar{Q}_{\rm sim}^{\rm Iso}$, as well as their standard deviations, are listed in Table \ref{tab:QandLimits}.  

It is clear that the estimates made assuming NFW and pseudo-isothermal profiles agree with each other within uncertainties. As stated, this is because the core radius is typically smaller than the scales of interest ($r_h$), and the phase-space density is calculated on scales where the NFW and pseudo-isothermal profiles are similar. Secondly, it is clear that the estimates based on VL2 are smaller than those from stellar kinematics. The differences are some factors of $\sim 2$--$5$, indicating $\eta_* \sim 1.3$--$1.7$. In the case of Leo IV it is as large as a factor of $\sim 9$, or $\eta_* \sim 2$.

Lower mass limits on the DW sterile neutrino are then derived. The NFW and pseudo-isothermal profiles yield very similar results, and in column 10 of Table \ref{tab:QandLimits} results for the pseudo-isothermal case are shown. The errors have been symmetrized conservatively such that both the upper and lower error-bars are enclosed if they are asymmetric. From these we determine the one-sided $95$\% C.L.~lower mass limits: $m_s^{\rm DW} \geq 2.5$ keV for Segue I, and the next strongest limit is Coma Berenices and Canes Venaviti II which are both $m_s^{\rm DW} \geq 1.5$ keV. 

\section{M 31 dSph count limits}\label{sec:counts}

The suppression of small-scale power due to DM streaming also manifests itself in the number of subhalos of massive halos. The suppression scale, and therefore the DM mass, can be constrained by comparing the subhalo distributions of suitable $N$-body simulations to observations of MW dSph \cite{Polisensky2010rw}. However, the census of MW dSphs suffers from significant radially-biased incompleteness \cite{Tollerud2008ze}, and being a single galaxy, the comparison must also take into account significant statistical and systematic uncertainties. 

The dSphs of our nearest neighbor, M 31, provide a compelling comparison set. In particular, we have the benefit of being outside the galaxy, yet close enough to detect dSphs down to fairly low luminosities. The Pan-Andromeda Archeological Survey (PAndAS) has complete coverage out to $\sim 150$ kpc from M 31 and sensitivity to dSphs down to luminosities of $\sim 10^5 L_\odot$ \cite{Richardson2011ru}. The dSph distributions were recently analyzed in Ref.~\cite{Yniguez2013qia}. They find that the M 31 dSph distributions are a better match to $\Lambda$CDM predictions than the MW dSph distributions, and argue for significant incompleteness of MW dSph under the $\Lambda$CDM paradigm. 

\begin{figure}
\includegraphics[width=3.25in]{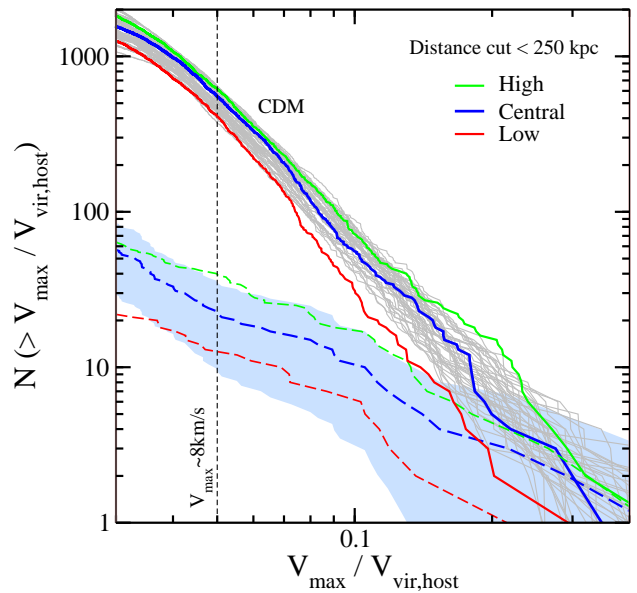}
\caption{Cumulative subhalo counts as functions of $V_{\rm max}$ normalized to the host virial velocity. The series of simulations labeled CDM are from the ELVIS suite of $\Lambda$CDM simulations of the Local Group; the three highlighted runs encompass the extremes (high and low in green and red) and the central (in blue). The dashed lines are the WDM analogues, all for $m_s^{\rm DW} = 6$ keV. The shaded region denotes the $2\sigma$ scatter in the central WDM estimated from CDM, which is consistent with the high and low WDM runs (see the text for details). \label{fig:distribution}}
\vspace{-1.\baselineskip}
\end{figure}

We derive $m_s^{\rm DW}$ limits based on comparisons of the observed M 31 dSph population to a series of WDM collisionless zoom-in simulations, requiring that the subhalo counts match or exceed the observed dSph counts. The details of the simulations are described in Ref.~\cite{horiuchi2013}; here, we summarize the relevant properties. All simulations were conducted with the $N$-body simulation code \texttt{GADGET-2} \cite{Springel2005mi} with WMAP7 parameters \cite{Larson2010gs}. We use initial conditions simulated as part of the ELVIS project, which is a suite of $48$ zoom-in simulations designed to study the Local Group \cite{ELVIS}. The suite consists of $24$ halos in paired systems that are chosen to resemble the MW and M 31 in mass and phase-space configuration, in addition to $24$ halos that are isolated mass-matched analogues. The halo mass varies between $1.0$--$2.8 \times 10^{12} M_\odot$, with associated virial radii $r_{\rm vir} = 263$--$370$ kpc and maximum circular velocities $V_{\rm max} = 155$--$225 \, {\rm km/s}$. To build the WDM initial conditions, CDM transfer functions generated using the \texttt{CAMB} code \cite{Lewis1999bs} for the ELVIS project were modified according to the analytic prescriptions of Ref.~\cite{Abazajian2005gj}, Eqs.~(10--12) to produce the WDM transfer functions. The simulations were run with identical mass resolution (particle mass $1.9 \times 10^5 M_\odot$). At this resolution, the simulations are complete to subhalos with $V_{\rm max} \gtrsim 8 \, {\rm km/s}$ \cite{ELVIS}. Halo substructure was identified with the Amiga Halo Finder \cite{Knollmann2009pb}. 

\begin{table}
\caption{Subhalo counts for CDM and WDM simulations with mass $m_s^{\rm DW}=3$, $6$, $10$, and $15.5$ keV (corresponding to thermal masses of $0.76$, $1.28$, $1.88$, and $2.62$ keV, respectively), with cuts of $8<V_{\rm max}/({\rm km/s})<60$ and various distance cuts are applied as indicated. The number of observed dSph of M 31 is also shown. \label{tab:counts}}
\begin{ruledtabular}
\begin{tabular}{lrrr}
Simulation		& $D<250\, {\rm kpc}$  	&  $D<200\, {\rm kpc}$	& $D<150\, {\rm kpc}$	\\
\hline 
Central CDM		& $454 \pm 122$		& $368 \pm 100$		& $246 \pm 62$		\\
Central $15.5$ keV	& $72 \pm 16.4$		& $53 \pm 13.2$		& $32 \pm 8.1$		\\
Central $10$ keV	& $37 \pm 9.3$			& $28 \pm 7.5$			& $13 \pm4.3 $		 \\
Central $6$ keV	& $19 \pm5.7$			& $13 \pm4.3$			& $8\pm3.0$		 \\
Central $3$ keV	& $6 \pm2.4$			& $5 \pm2.2$			& $4\pm2.0$		 \\
\hline 
Observed 			& $>29$				& $>26$				& $18$
\end{tabular}
\end{ruledtabular}
\end{table}

The subhalo distributions, as functions of $V_{\rm max}/V_{\rm vir,host}$, where $V_{\rm vir,host}$ is the host halo virial velocity, are shown in Figure \ref{fig:distribution}. A distance cut of $<250$ kpc is applied. The resolution limit $V_{\rm max} > 8 \, {\rm km/s}$ is plotted for the mean $V_{\rm vir}$ of $ \approx 263$ km/s for clarity. By normalizing by $V_{\rm vir,host}$, some of the scatter in subhalo distributions is reduced, but still captures the more dominant and important scatter in the total number of subhalos which, as described below, is used to obtain limits on sterile neutrinos. The $48$ ELVIS simulations are shown in grey and provide a measure of the scatter in predictions for a single $\Lambda$CDM cosmology, i.e., the combined systematic uncertainty due to cosmic variance, range of plausible M 31 halo mass, and low number Poisson scatter. We take a flat prior on the host halo mass; thus the scatter is a conservative overestimate of the true scatter due to the uncertainty of the mass of M 31. The overall scatter is non-neglible, being some factors of $\sim 2$ at the minimum and increasing as the number of subhalos decrease. 

In order to test whether the scatter in CDM changes in WDM, we adopt three runs from ELVIS capturing the two extremes (indicated by the green and red lines) and the central behavior (indicated by the blue line). These are simulated for $m_s^{\rm DW} = 6$ keV, shown by the dashed lines. To compute the subhalo scatter for CDM, we calculate the mean $\bar{N}_{\rm subs}^{\rm CDM}$ and standard deviation $\sigma_{\rm subs}^{\rm CDM}$ of the cumulative subhalo counts according to the ELVIS suite, and determine the normalized standard deviation, $\hat{\sigma}=\sigma_{\rm subs}^{\rm CDM}/\bar{N}_{\rm subs}^{\rm CDM}$, as a function of $N_{\rm subs}$. We then apply this distribution according to the number of subhalos in WDM simulations, i.e., $\hat{\sigma}(N_{\rm subs}=N_{\rm subs}^{\rm WDM})$. The blue shaded band in Figure \ref{fig:distribution} shows the $2\sigma$ region about the Central-6keV run estimated by this method. The inclusion of the two extreme WDM simulations within this band demonstrates the applicability of this method. 

\begin{figure}
\includegraphics[width=3.25in]{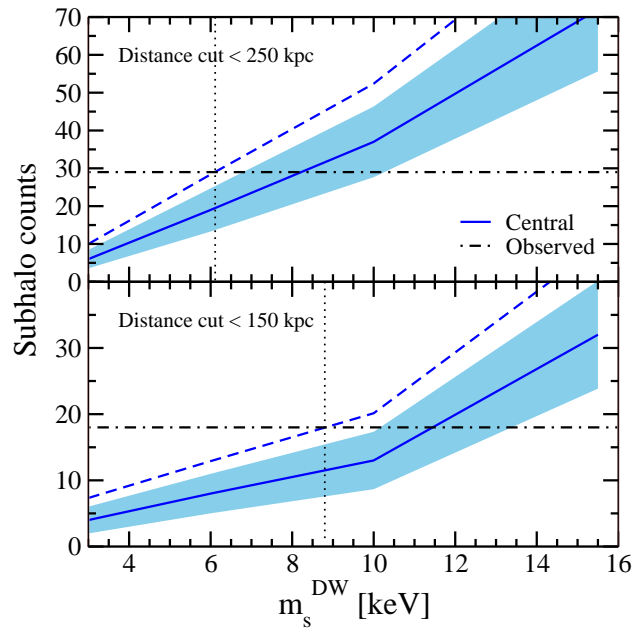}
\caption{The number of subhalos seen in simulations (solid) and its $1\sigma$ uncertainty (shaded), compared to the number of dSphs of M 31 (dot-dashed line), for distance cuts of $250$ kpc (top panel) and $150$ kpc (bottom panel). The dashed line denotes the one-sided $95$\% uncertainty and the vertical dotted lines indicate the resulting one-sided $95$\% C.L.~mass limit. \label{fig:countsvsmass} }
\vspace{-1.\baselineskip}
\end{figure}

Having established the validity of applying the CDM scatter to WDM, we simulate the central initial condition for $m_s^{\rm DW}$ masses of $3$, $6$, $10$, and $15.5$ keV (equivalent to thermal masses of $0.76$, $1.28$, $1.88$, and $2.62$ keV, respectively). The subhalo counts using cuts of $8 \, {\rm km/s}<V_{\rm max}<60 \, {\rm km/s}$ and various distance cuts are summarized in Table \ref{tab:counts}. We compare these to the observed M 31 dSph counts, which are derived mainly from Ref.~\cite{McConnachie2012vd}, with the addition of recently discovered And XXX \cite{Conn2012um}, And XXXI, and And XXXII \cite{Martin2013cha}. The vast majority of these have estimated $V_{\rm max}$ greater than $8$ km/s \cite{Tollerud2011mi}. We exclude from this list M 33, since our focus is on dwarf galaxies, consistent with our upper $V_{\rm max}$ cut. 

Unsurprisingly, our smallest distance cut $<150$ kpc, which is consistent with the completeness range of M 31 dSph, provides the strongest constraint: at $1 \sigma$ deviation our $10$ keV run has barely enough subhalos to match the observed number of M 31 dSphs. The interpolation with mass is shown in Figure \ref{fig:countsvsmass}, and shows that at $m_s^{\rm DW} = 8.8$ keV the one-side $95$\% scatter matches the (minimum) required number of M 31 dSph; we quote this mass as our one-sided $95$\% C.L.~lower mass limit. The limits for the other distance cuts are $6.1$ keV and $7.2$ keV for $<250$ kpc and $<200$ kpc, respectively.

\section{M 31 X-ray limits}\label{sec:xray}
To obtain improved X-ray constraints on the mixing angle ($\theta$) between sterile and active neutrinos, we take advantage of significantly deeper \chandra\ ACIS data than has been used in previous work \cite{watson12a}. The central part of the galaxy, where the signal is expected to be strongest, has been repeatedly imaged as part of monitoring programs, enabling a very deep composite image. This is optimal for the detection and removal of the faint point sources that constitute most of the X-ray emission from \src. In this respect, \chandra\ is far superior to \xmm\ or \suzaku.

\begin{figure}
\includegraphics[width=3.25in]{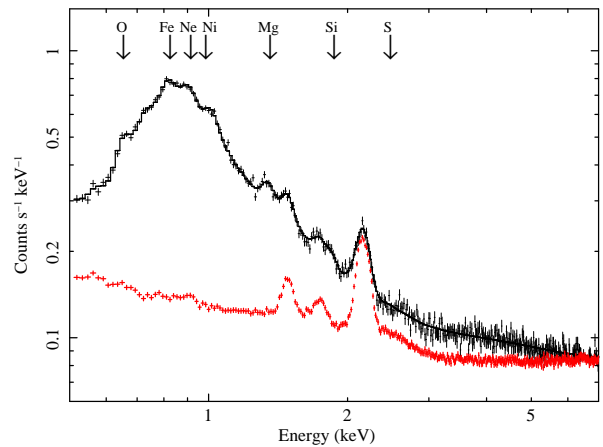}
\caption{Coadded on-axis ACIS-I spectral data (black points), without 
background subtraction. In red we show an
estimate of the background extracted from identical regions of blank sky 
fields. The solid black line is the best-fitting model, folded through
the instrumental response, which
fits the data well. We note the location of astrophysically important
emission lines expected for the collisionally ionized plasma present in the
bulge of \src.\label{fig_spectrum}}
\vspace{-1.\baselineskip}
\end{figure}

We assembled all 70 \chandra\ datasets taken in the ACIS-I configuration that were centered on the nucleus of \src\ and were publicly available as of Mar 06 2013. We used ACIS-I due to its lower, more stable instrumental background and larger unobstructed field of view than ACIS-S. Each dataset was reduced and processed with the CIAO 4.3 software suite following standard procedures, and the astrometry was corrected onto a standard reference frame by matching bright sources detected in each observation, as described in \cite{humphrey08a}. Images and spectra were co-added to produce a total on-axis exposure of 267~ks. Source detection was performed on the co-added image with the {\em wavdetect} CIAO tool, and data within three times the $90$\%\ encircled energy ellipses estimated for each source were excluded from subsequent analysis. A composite spectrum, and count-weighted response matrices, were extracted from the exposed regions of the CCDs, excluding the central 2\arcmin\ and any detected sources. In Figure \ref{fig_spectrum}, we show the spectral data, along with the equivalent data extracted from identical regions of the standard ``blank sky field'' events files, which provides a rough estimate of the background. These data were supplemented by 17 additional ACIS-I datasets that were offset from the centre of the galaxy by $\sim$1--5\arcmin. These data were analysed separately, and their spectra added, for a total exposure of 137~ks. The effective area curves were averaged, weighting by the relative expected line flux. 

Following \cite{humphrey11a}, we fitted the spectra with a physically motivated model using {\em vers.} 12.7.1 of the {\tt XSPEC} spectral fitting package \cite{xspec}. The data were mildly rebinned (to ensure $>$20 photons per bin), which aids in error bar computation, and we minimized the Cash-C goodness-of-fit statistic \cite{humphrey09b}. The model comprised a powerlaw and two thermal (APEC, \cite{apec}) plasma components to account for the cosmic and Galactic X-ray background, plus two broken power law models (not multiplied by the effective area) and three Gaussian lines to account for the instrumental background. To account for emission from \src, we included a power law and two APEC plasma components (kT=$0.32$ and $0.78$keV) that were modified by photoelectric absorption with the nominal hydrogen column density for the centre of \src\ ($1.27\times 10^{21}cm^{-2}$: \cite{kalberla05a}). We allowed the abundance of Fe, N, O, Ne, Mg, Si and Ni to fit freely. Other elements were tied to Fe in their Solar ratios (\cite{asplund04a}), except He, which was fixed at its Solar value. This gives a satisfactory fit to the spectrum (Fig~\ref{fig_spectrum}). The temperature of the softer component is close to that inferred from \xmm\ observations of \src\ \cite{liu10a}, while the hotter component and powerlaw are expected to be a good parameterization for residual,  unresolved sources \cite{revnivtsev08a}.

All of the obvious emission lines in the spectrum are well-fitted by our model, since they are coincident either with astrophysically interesting lines expected from a thermal plasma or with the instrumental fluorescent features (Fig~\ref{fig_spectrum}). To obtain upper limits on \mixing, we therefore added an additional, narrow, photo-absorbed Gaussian line to the spectral model at energy $E_\gamma=m_s/2$ and with flux ($F_\gamma$) given, after \cite{boyarsky08a,watson12a}, by:
\begin{equation}
F_\gamma = 10^{-7} \, {\rm erg\ s^{-1}\ cm^{-2}} \times \left(\frac{M_{DM}^{FOV}}{10^{11}M_\odot}\right) 
D^{-2} m_s^5 \sin^2 2\theta 
\end{equation}
where $M_{DM}^{FOV}$ is the projected mass in the field of view of the observation, $D$ is the distance in Mpc (for which we adopt $0.784$ Mpc). We estimated \mdm\ ($1.6\times 10^{10}$\msun\ for the on-axis spectrum) by integrating the DM surface density, estimated from the model of \cite{seigar08a}, over the field of view of each individual pointing. We then appropriately averaged each value to ensure the correct line count-rate in the composite spectra. 

To determine an upper limit on \mixing\ for a given \mparticle, the line (at fixed energy) was added simultaneously to the on-axis and offset spectra, and its normalization varied (while fitting all other parameters) until the fit statistic increased by $4.61$, corresponding to a 95\%\ confidence interval for two parameters of interest. This approach is similar to the ``statistical'' method of \cite{boyarsky08a}, although we have appropriately included the required statistical uncertainties on the background model. In Fig~\ref{fig:results}, we show our measured upper limits on \mixing. Because the fluxes of the astrophysical and instrumental lines are not known {\em a priori}, they are degenerate with any coincident sterile neutrino decay line. This reduction in sensitivity is immediately apparent in the jagged upper limit curve. A major source of uncertainty in this measurement is the precise value of \mdm \cite{boyarsky08a}. For example, if we use the DM profile model $C_1$ of Ref.~\cite{klypin02a}, \mdm\ is increased by $\sim$15\%\ in the core, resulting in correspondingly tighter constraints on \mixing.

\section{Discussion}\label{sec:conclusion}
\begin{figure}[t]
\includegraphics[width=3.25in]{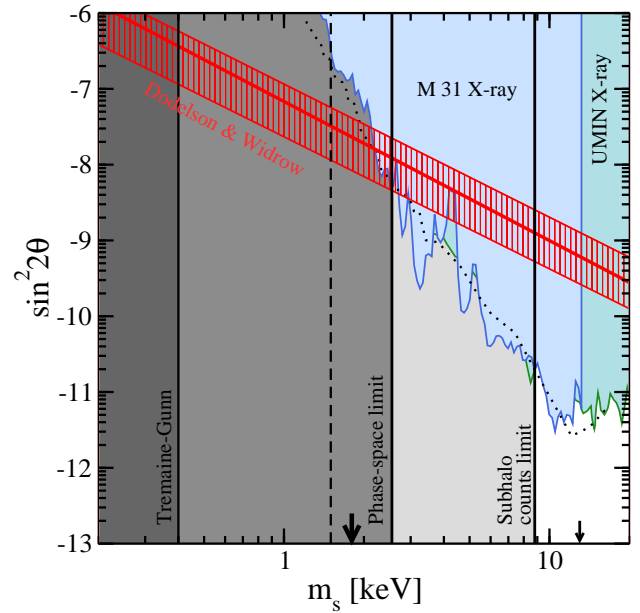}
\caption{Constraints on sterile neutrino parameters. Shaded areas are excluded regions: 95\% C.L.~upper limits derived from the X-ray modeling of \src\ (labeled ``M31 X-ray''), the results from Ref.~\cite{watson12a} shown for comparison (dotted; see text), and upper limits from Suzaku observations of Ursa Minor \cite{Loewenstein2008yi} (labeled ``UMIN X-ray''); vertical lines show lower mass limits from Tremain-Gunn phase-space considerations ($m_s \sim 0.4$ keV) \cite{Tremaine1979we}, Coma Berenices phase-space ($m_s^{\rm DW} \sim 1.5$ keV, dashed line), Segue I phase-space ($m_s^{\rm DW} \sim 2.5$ keV), and M 31 subhalo counts ($m_s^{\rm DW} \sim 8.8$ keV). The big and small arrows on the abscissa indicate lower limits from Ref.~\cite{Boyarsky2008ju} and Ref.~\cite{Polisensky2010rw}, respectively. The DW sterile neutrino model of Ref.~\cite{Dodelson1993je} and its associated upper and lower bounds \cite{Asaka2006nq} are shown and labeled.
\label{fig:results}}
\vspace{-1.\baselineskip}
\end{figure}

The one-sided $95$\% C.L.~lower and upper limits from the Local Group are shown in Figure \ref{fig:results}. These include lower limits from phase-space arguments of MW dSphs ($m_s^{\rm DW} \gtrsim 2.5$ keV), lower limits from subhalo counting comparison to M 31 dSphs ($m_s^{\rm DW} \gtrsim 8.8$ keV), and upper limits based on X-ray observations of M 31. Combined, these decisively constrain the canonical Dodelson-Widrow (DW) production mechanism for generating sufficient sterile neutrinos to match the DM abundance at $>99$\% C.L. 

Phase-space arguments have been argued to be among the most robust methods to constrain WDM, but they have not been strong enough to rule out the DM sterile neutrino when coupled with X-ray limits \cite{Boyarsky2008ju} (indicated by the larger arrow in Figure \ref{fig:results} at $1.8$ keV). Our newly added Segue I dSph, combined with updated X-ray limits based on deep Chandra observations of M 31, excludes the entire DW model parameter space, including the wider range due to hadronic model uncertainties \cite{Asaka2006nq} (red hatched), at $95$\% C.L. The exception is around $m_s^{\rm DW} \approx 4.3$ keV, where a strong X-ray background line in the M 31 data prevents a strong limit on a sterile neutrino decay line. However, limits from Suzaku---with vastly different backgrounds and in particular weaker lines---already exclude this region \cite{Loewenstein2008yi}, as shown in Figure \ref{fig:results}. If Segue I is not included, the mass limit is weakened to $1.5$ keV (dashed vertical line) and allows a DW sterile neutrino of $m_s^{\rm DW} \approx 2$ keV to generate the observed cosmological DM abundance. However, including limits from subhalo counting, all of the DW parameter region is comfortably excluded at $>99$\% C.L. 

For the same dwarfs, our limits are weaker than those of Ref.~\cite{Gorbunov2008ka}, where the authors adopted significantly higher phase-space density estimates (e.g., $5\times10^{-3} ({\rm M_\odot/ pc^3 })({\rm km/s})^{-3}$ for Leo IV and Canes Venatici II). These follow from Ref.~\cite{Simon2007dq}, where the {\em central} density is used to estimate $Q$, as opposed to our conservative estimate based on the mean density within $r_h$. Also, the stellar velocity dispersion is assumed in that work to be the same as the DM velocity dispersion ($\eta_*=1$). For these reasons, we obtain weaker but more robust limits. Our limits are similar in numerical value to those of Ref.~\cite{Boyarsky2008ju}, where the authors assume $\eta_*=1$ but consider the phase volume defined by the escape velocity of DM particles. 

Our limits from subhalo count are somewhat weaker than previous constraints placed using MW dSphs \cite{Polisensky2010rw} (indicated by the smaller arrow in Figure \ref{fig:results} at $13.3$ keV). However, bearing in mind that these MW limits rely on corrections of factors of $2$--$4$ for missing dSphs---for example, in their most constraining distance bin ($< 50$ kpc from the MW), the correction is from $7$ to $16$ dSphs---our results compare quite favorably. Part of the reason is the different method used to obtain the limit. We take the Bayesian approach such that: given an observed number of dSphs, then what is the probability that a model with mass $m_s^{\rm DW}$ could produce the observation. On the other hand, Ref.~\cite{Polisensky2010rw} included fluctuations in both the model and the observed number of dSph to set their limits, which weakens their limits. Our limits are stronger than those of the MW without incompleteness corrections ($7$ keV) \cite{Lovell2013ola}.

For X-ray limits, despite the significantly deeper data used in our analysis ($\sim$400~ks versus 50~ks), the limits of Ref.~\cite{watson12a} are tighter for some range of \mparticle. This is particularly true for $m_s \sim4.3$~keV, corresponding to $E_\gamma \sim 2.1$~keV, which is coincident with a strong background line. In practice, Watson et al.\ \cite{watson12a} adopted the value of \mixing\ for which the line flux equalled the background-subtracted flux at each energy. This, however, requires the background to be known exactly, whereas we explicitly included background uncertainties in our measurements, which most likely accounts for the differences. 

Although we disfavor the DW mechanism as the sole production of DM, sterile neutrinos may be generated by resonant oscillations or non-oscillation channels. These result in ``mixed'' DM consisting of a warm (non-resonant production) and a colder (the resonant or non-oscillation production) component. They are not as constrained by our limits \cite{Boyarsky2008mt,Boyarsky2008xj}. For example, resonant oscillation allows for a smaller mixing to generate the required DM abundance, which helps evade the X-ray constraints. Furthermore, the velocity dispersion is colder, meaning $q_{\rm max}$ is larger than for DW, relaxing the phase-density limits. For example, we estimate that in the mixed models of Ref.~\cite{Boyarsky2008xj}, a $3$ keV sterile neutrino generating the required DM abundance in lepton asymmetries of $L_6=10^6(n_{\nu_e} - n_{\bar{\nu}_e})/s = 10$ (16, 25) results in primordial $q_{\rm max}^{\rm res} \approx 35$ (160, 30). These are significantly larger than the fine-grained phase-space maximum for the DW model, $q_{\rm max}^{\rm DW} \approx 14$ [all in units $10^{-5} ({\rm M_\odot/ pc^3 })({\rm km/s})^{-3}$]. In non-oscillation production mechanisms, the mixing angle may be arbitrarily small and the velocity dispersion is also colder \cite{Kusenko2006rh,Shaposhnikov2006xi,Merle2013wta}. 

WDM models have been investigated as attractive solutions to many of the challenges faced by CDM on sub-galactic scales (see, e.g., Ref.~\cite{Weinberg2013aya} for a recent coverage of key issues). One recent example is the ``Too big to Fail'' problem (TBTF) \cite{BoylanKolchin2011de,BoylanKolchin2011dk}. Various WDM models have been investigated in the literature, including WDM of thermal particle masses $m_{\rm WDM} = 1$--$4$ keV \cite{Anderhalden2012jc,Lovell2013ola,Polisensky2013ppa}, and mixed WDM models with mass $m_{\rm WDM} = 2$ keV and smaller \cite{Lovell2011rd,Anderhalden2012jc}. These studies find that masses of $m_{\rm WDM} =1$--$2$ keV are required to solve TBTF; above $2$ keV, there is insufficient difference from CDM in the subhalo kinematics \cite{Schneider2013wwa}. Hence, the lower end of the solution mass range is inconsistent with our limits, leaving only a narrow range of possible mass. We caution that the physically relevant quantity for a detailed comparison is the cutoff scale and shape; the mass alone is insufficient since a mixed model of a given mass has a more diluted cutoff due to the cold component than a sterile neutrino of the same mass. 

Our conclusion, while similar to those discussed for some Lyman-$\alpha$ and MW dSphs abundance matching limits \cite{Boyarsky2008xj,Polisensky2010rw}, are independent and most importantly robust, making them decisive on whether the DW mechanism can generate the entire DM abundance. Future dSph discoveries are expected by upcoming surveys, which will enable stronger limits that go deeper into the parameter space of mixed sterile neutrino models. 

\medskip
{\bf Acknowledgments.---}
We thank Marc Seigar for providing his mass profile data in electronic form. We thank John Beacom, James Bullock, Enectali Figueroa-Feliciano, Shea Garrison-Kimmel, Evan Kirby, and Alexander Kusenko for useful discussions. This work is supported by a JSPS fellowship for research abroad (SH), a Gary McCue Fellowship offered through UC Irvine (PJH), and NSF Career Grant PHY-1159224 (KNA). JO thanks the financial support of the Fulbright/MICINN Program.

\bibliography{M31.bib}

\end{document}